# One-way helical electromagnetic wave propagation supported by magnetized plasma

Biao Yang, Mark Lawrence, Wenlong Gao, Qinghua Guo, Shuang Zhang*

School of Physics & Astronomy

University of Birmingham

Birmingham, B15 2TT, UK

*s.zhang@bham.ac.uk

**Abstract:** In this paper we reveal the presence of photonic one-way helical surface states in a simple natural system - magnetized plasma. The application of an external magnetic field to a bulk plasma body not only breaks time-reversal-symmetry but also leads to separation of Equi-Frequency Contour surfaces (EFCs) to form topologically nontrivial gaps in k space. Interestingly, these EFCs support topologically protected surface states. We numerically investigate an interface between magnetized plasma, using a realistic model for parameter dispersion, and vacuum, to confirm the existence of one-way scatter-immune helical surface states. Unlike previous proposals for achieving photonic one-way propagation, our scheme does not require the use of artificial structures and should therefore be simple to implement experimentally.

The pursuit of one-way scatter-immune transportation of light has recently become a hot research topic not least because of the clear technological benefits of being able to manipulate electromagnetic waves while maintaining perfect transmission. Its apparent carrier is surface wave, by definition, is a wave bounded by the interface between two semi-infinite media. Traditional surface wave usually endures non-ignorable scattering loss when encountering any interfacial disorders, such as sharp bend. Recently, many kinds of photonic nontrivial surface wave, which show the robust property of one-way scatter-immune transportation, have been proposed by means of analogue of electronic topological insulator [1,2]. The defining feature of topological phases is bulk-edge correspondence. Bulk-edge correspondence, referring to the surprising dependence of boundary excitations on the characteristics of the bulk propagating modes rather than the local properties of the boundary, rebuilds our concepts on energy band theory. After Haldane proposed the photonic analogue of Quantum Hall Effect in 2008 [3,4], topological photonic edge states have been investigated both theoretically [5-10] and experimentally [11-15]. Simultaneously, magnetized near-zero-epsilon metamaterials [16] was also proposed to provide one-way photonic states in 2-D plane. Obviously, a wholly new field - topological photonics [17] has emerged to support metamaterials. However, what cannot be neglected is that fabrication and assembly of metamaterials is still challenging and time consuming. Here, we propose a simple natural system - magnetized plasma to realize helical one-way surface propagation. More importantly, the topological transportation of electromagnetic surface waves can be reconfigurable by simply adjusting the external magnetic field or the plasma density.

Magnetized plasma has been recently investigated to show interesting electromagnetic properties such as sub-diffractional imaging [18] and magnetic field induced transparency [19]. Under a strong magnetic field, the movement of the free electrons in the cold plasma is confined in the plane perpendicular to the applied magnetic field. On the other hand, the electrons can move freely along the direction of the magnetic field. It has been shown previously that with extremely strong magnetic field [18], the wave can propagate almost diffractionlessly along the direction of the magnetic field, in a similar way as wave propagate inside a hyperbolic metamaterials [20], which currently is a very active research topic including subwavelength imaging [21-23], negative refraction [24], spontaneous emission enhancement with a large Purcell Factor [25] and especially topological ordered metamaterials [26].

For propagation of electromagnetic wave with angular frequency $\omega$ in the lossless plasma, the lossless plasma behaves as a free-electron model, whose electromagnetic response can be described by the Drude permittivity $\varepsilon = 1 - \omega_p^2/\omega^2$, where $\omega_p = \sqrt{Ne^2/\varepsilon_0 m}$ is the plasma frequency, $N$ being the electron concentration, $e$ and $m$ denote elementary electron charge and electron mass, respectively. If a magnetic field is aligned in the z direction, the electromagnetic response of the lossless magnetized plasma can be described by the following local, homogeneous permittivity tensor [18,27,28]

$$\vec{\vec{\varepsilon}} = \begin{bmatrix} \alpha & i\delta & 0 \\ -i\delta & \alpha & 0 \\ 0 & 0 & \beta \end{bmatrix}, \qquad [1]$$

where, $\alpha = 1 - \omega_p^2/(\omega^2 - \omega_B^2)$, $\delta = -\omega_B \omega_p^2/\omega(\omega^2 - \omega_B^2)$ and $\beta = 1 - \omega_p^2/\omega^2$. The cyclotron frequency $\omega_B = eB/m$ is determined by the applied static magnetic field $\vec{B}$.

Considering a wave propagating along z direction with circular polarizations as base states, this relative permittivity of the magnetized plasma can be reduced to a diagonal tensor $\mathrm{Diag}[\varepsilon_R, \varepsilon_L, \beta]$, where $\varepsilon_{R,L} = 1 - \omega_P^2/[\omega(\omega \pm \omega_B)]$, '+,−' correspond to right and left polarized states, respectively. To meet our hyperbolic demand in z direction which is parallel to the static magnetic field, the operating frequency must satisfy $\omega < \omega_p$ in the lossless condition. Whereas the applied magnetic field draws off-diagonal components into the relative permittivity, magnetized plasma shows two bandgaps in k-space with condition of $\varepsilon_{L,R} > 0$. Thus, the operating angle frequency has to be less than $\omega_B$ and satisfies $\omega_P^2 < \omega(\omega + \omega_B)$. Based on these constraints, we can find $\varepsilon_R < 1 < \varepsilon_L$, which means EFCs of vacuum is located in the EFCs' bandgaps of the magnetized plasma, as shown in Fig. 1(a). It is shown that these two materials (magnetized plasma and vacuum) have overlapped forbidden bands, which is an essential condition to realize one-way scatter-immune transportation. Fig. 1(b) shows corresponding polarization properties and bandgaps of magnetized plasma. The EFCs embodies circular polarization states and linear polarization states at pole points and infinity circle, respectively.

We now turn to studying how these topological features manifest themselves on the boundary of magnetized plasma. In what follows we investigate systems with continuous translational invariance in the z direction, thereby conserving $k_z$. Here, by assuming the surface wave exponentially decays along either direction from y-z plane (i.e., the half space $x>0$ is occupied by isotropic media such as vacuum, whereas the magnetized plasma is located in $x<0$), we use the method proposed by Dyakonov in 1988 [29] to calculate these nontrivial surface wave. On the vacuum side, there are two orthogonal eigen modes which can be expressed as,

$$\begin{bmatrix} E_1 & H_1 \\ E_2 & H_2 \end{bmatrix} = \begin{bmatrix} q & 0 & -k_{vac} & 0 & 1 & 0 \\ 0 & 1 & 0 & -q & 0 & k_{vac} \end{bmatrix},$$

[2]

where $k_{vac} = i\sqrt{q^2-1}$ is imaginary representing decay constant along positive $x$ direction and $q$ is the absolute value of in-plane propagation wave vector. Likewise, we can write down the decay constant on magnetized plasma side $-ik_{m-p}$ as a function of vector $\vec{q}$ (instead of $q$ due to gyrotropic property of magnetized plasma). Two independent eigen modes of magnetized plasma can be obtained through solving Maxwell's Equations,

$$\begin{bmatrix} E_3 & H_3 \\ E_4 & H_4 \end{bmatrix} = \begin{bmatrix} E_{3x} & E_{3y} & E_{3z} & H_{3x} & H_{3y} & H_{3z} \\ E_{4x} & E_{4y} & E_{4z} & H_{4x} & H_{4y} & H_{4z} \end{bmatrix},$$

[3]

which are functions of $\vec{q}$. To this end, we have obtained four eigen modes that are localized on both sides of the surface. The tangential components of these fields are continuous across the interface, leading to the determinant problem of a $4 \times 4$ constraint matrix,

$$\det \begin{bmatrix} 0 & 1 & E_{3y} & E_{4y} \\ -k_{vac} & 0 & E_{3z} & E_{4z} \\ 1 & 0 & H_{3y} & H_{4y} \\ 0 & k_{vac} & H_{3z} & H_{4z} \end{bmatrix} = 0.$$

[4]

The results of the above equation are shown in Fig. 2(a) and (c), which reveal that in the gap of EFCs any given surface can support just one propagating mode. This means that the spatial separation of left and right moving surface waves at certain $k_z$ prevents the occurrence of backscattering from any z-invariant disorder as shown in Fig. 2(b).

Interestingly, after zooming in the point 'C' showed in Fig. 2(c) where the surface state is very close to the EFC of vacuum, it is discovered that the surface state is not merging into the EFC of vacuum, rather, it conformally bends around the EFC of vacuum and connects to the other branch of the EFC of magnetized plasma. This indicates that even at $k_z = 0$ $\vec{q} = k_y \hat{y}$, there exists a single unidirectional surface state on each surface. The dispersion relation at $k_z = 0$ can be expressed as [16],

$$\mp \frac{q\delta}{\alpha \varepsilon_f} + \frac{\sqrt{q^2 - \varepsilon_f}}{\varepsilon_f} + \frac{\sqrt{q^2 - \varepsilon\mu}}{\varepsilon} = 0 \qquad [5]$$

where $\varepsilon_f = \alpha^2 - \delta^2 / \alpha$ and $\mp$ represents to top and bottom surface, respectively. After substituting the realistic parameters of magnetized plasma into Eq. 5, we obtain the small difference between the wave vectors of the surface state and the vacuum bulk state at point 'C' as $\Delta k_y = 0.342$. However, since the surface states do not fall in the gap, they are not protected against z invariant defects.

This immunity to backscattering has also been confirmed using full wave simulations shown in Fig. 2(d) in which a right moving surface wave propagates seamlessly around a sharp defect. The simulation is performed in the x-y plane for three different propagation constants $k_z$ in the shadow area as indicated by 'A', 'B' and 'C' in Fig. 2(c). For point 'A' and 'C', since they are not located in the gap region, the electromagnetic wave can be scattered into the bulk states by z-invariant scatterers. On the other hand, at point 'B' where $k_z$ is in the middle of the gap, the surface wave is immune from scattering by the sharp edges, and therefore the numerical simulation confirms that it is a topologically nontrivial surface state. As expected, when the direction of the magnetic field is flipped, the propagation of the surface wave is also switched to the opposite direction. This may enable topological surface states with dynamically reconfigurable properties.

In conclusion, we have theoretically investigated the existence of one-way propagating surface wave between vacuum and magnetized plasma. The simulation confirms the presence of unidirectional backscattering-immune propagation of surface wave based on the simple natural system-magnetized plasma. Although the study has focused on free space magnetized plasma for manipulation of electromagnetic waves at microwave regime, it can be extended to terahertz regime by working with semiconductors with very small effect mass, such as InSb. Thus, this nontrivial edge states may also be observed in THz band by using magnetized semiconductors [30].

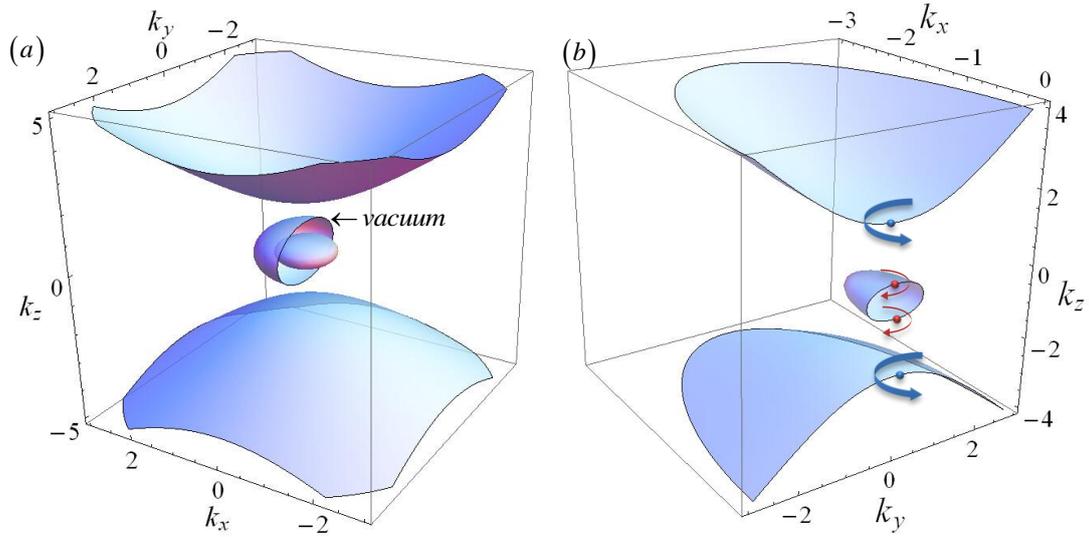

Fig. 1 EFCs and corresponding electric field polarization states of magnetized plasma. (a) The magnetized hyperbolic permittivity and EFCs, in which the ordinary mode is wrapped by vacuum (indicated). However, there still is a gap between the EFCs of the extraordinary mode and vacuum. (b) Polarization of electric field at pole points. At the minimal point of upper pseudo hyperbolic branch, light propagates with strict left circular polarization states. However, by $k_z$ approaching to infinity, the circular polarization states will verge to linear. The peak point of upper ordinary branch shows strict right circular polarization which is opposite to the minimal point, because these two states are lifted from a degeneracy point during nontrivial transformation from normal hyperbolic metamaterials. In the negative $k_z$, polarization presents opposite properties. In the plane of $k_z = 0$, the z component of electric field is zero again, but its polarization is slightly elliptical compared with those pole locations. The permittivity is calculated with respect to $\omega_p = 1.5\omega_0$, $\omega_B = 1.8\omega_0$. In this letter, all of wave vectors are measured with unit of $k_0 = \omega_0/c$.

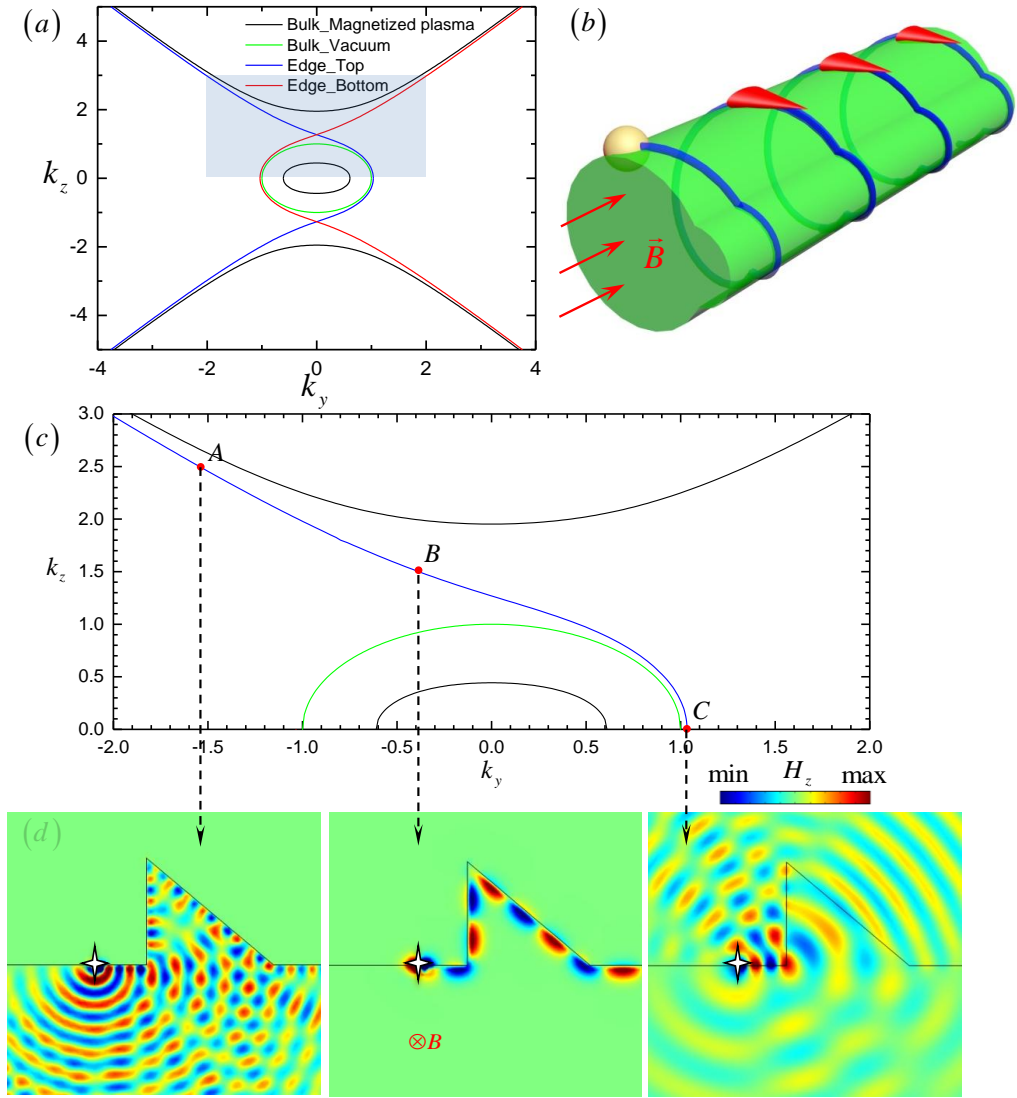

Fig. 2 Topologically protected surface states at the interface between a magnetic plasma and vacuum. (a) Bulk states and surface states of this system. Top and bottom relations are defined relatively. (b) Helical one-way back scattering-immune propagation in 3D. Chiral surface state propagating around the magnetized plasma with added cylindrical shape surrounded by air, despite the existence of raised cylinder back-scattering is forbidden due to the absence of anticlockwise modes. (c) Magnified shadow area of (a). (d) Field distribution simulated by commercial Comsol RF module. The parameter of the magnetized plasma are $\omega_p = 2\pi \times 8.976 GHz$, $\omega_B = 2\pi \times 10.771 GHz$ $B = 0.385T$ and the operating frequency $\omega_0 = 2\pi \times 5.984 GHz$. Since the collision frequency is several orders of magnitude lower than the operating frequency, this realistic case can still be regard as lossless.[18]